\documentclass[twocolumn,showpacs,amsmath,amssymb,prl]{revtex4}
\usepackage{graphicx}% Include figure files
\usepackage{dcolumn}% Align table columns on decimal point
\usepackage{bm}% bold math
\usepackage{subfigure}
\begin{document}

%\preprint{submitted to Phys. Rev. B}

\title { The $\gamma \rightarrow \alpha$ phase transition in cerium is not
isostructural }

\author {A. V. Tsvyashchenko$^{1,2}$, A. V. Nikolaev$^{2,5}$, A. I. Velichkov$^3$, A. V. Salamatin$^3$,
 L.~N.~Fomicheva$^1$, G. K. Ryasny$^2$, A. A. Sorokin$^2$, O. I. Kochetov$^3$,
and M. Budzynski$^4$}

\affiliation{$^1$Vereshchagin Institute for High Pressure Physics
of RAS, 142190, Troitsk, Russia}

\affiliation{$^2$Skobeltsyn Institute of Nuclear Physics, Moscow
State University, Vorob'evy Gory 1/2, 119991, Moscow, Russia}

\affiliation{$^3$Joint Institute for Nuclear Research, PO Box 79,
Moscow, Russia}

\affiliation{$^4$Institute of Physics, M. Curie-Sklodowska
University, 20-031 Lublin, Poland}

\affiliation{$^5$Institute of Physical Chemistry and
Electrochemistry of RAS, Leninskii pr. 31, 119991, Moscow, Russia}

\date{\today}

%--------------- ABSTRACT ---------------
\begin{abstract}
Using a small amount of radioactive $^{111}$Cd atoms implanted in cerium lattice to probe the hyperfine electric quadrupole interaction, we have detected a quadrupole electron charge density component in $\alpha$-Ce, which is absent in $\gamma$-Ce. The appearance of the quadrupole density in $\alpha$-Ce predicted by the quadrupole scenario of the phase transition demonstrates that the spacial symmetry is lowered at the $\gamma \rightarrow \alpha$ phase transition. This finding makes cerium the first element where the symmetry change is driven exclusively by the valence electron degrees of freedom while the atomic centers of mass (cerium nuclei) occupy the face centered cubic positions. We also report on nuclear quadrupole interactions in other high pressure phases of cerium: $\alpha''$ ($C2/m$ space symmetry) and $\alpha'$ ($\alpha$-U structure).
\end{abstract}


\maketitle

The crystal space group symmetry is the most important characteristic, which is used for the identification of a macroscopic collection of atoms or molecules. A transition to another allotropic form of an element is therefore fully described by the change of its space symmetry. This fact is exploited in the Landau theory of phase transitions \cite{Lan1}, but the consideration fails at the apparently ``isostructural" $\gamma \rightarrow \alpha$ phase transition in cerium \cite{Bri} occurring at a pressure of 0.7 GPa at room temperature and accompanied by a ~16\% volume collapse \cite{Law,Kos}. To distinguish the essentials of these twin phases one has to introduce a very important characteristic equal in its significance to the crystal space symmetry. Since the establishing of the face centred cubic (fcc) symmetry of both phases in 1949 \cite{Law} the transition has become a battleground of various microscopic theoretical models offering different parameters controlling the phase transition. Two models - a Kondo volume collapse \cite{All,Lav} and a Mott-like transition for 4$f$ electrons
\cite{Joh} - are well represented in the literature \cite{Lip}. Despite much effort, there is still no consensus on the issue \cite{Joh1}. Needless to say, these models take the isostructural nature of the phase change as a well established fact.

\begin{figure}
%\vspace{1mm} \resizebox{0.25\textwidth}{!} {
% \includegraphics{fig1_TDPAC.eps}   }

\subfigure{\resizebox{3.5cm}{!}{\includegraphics{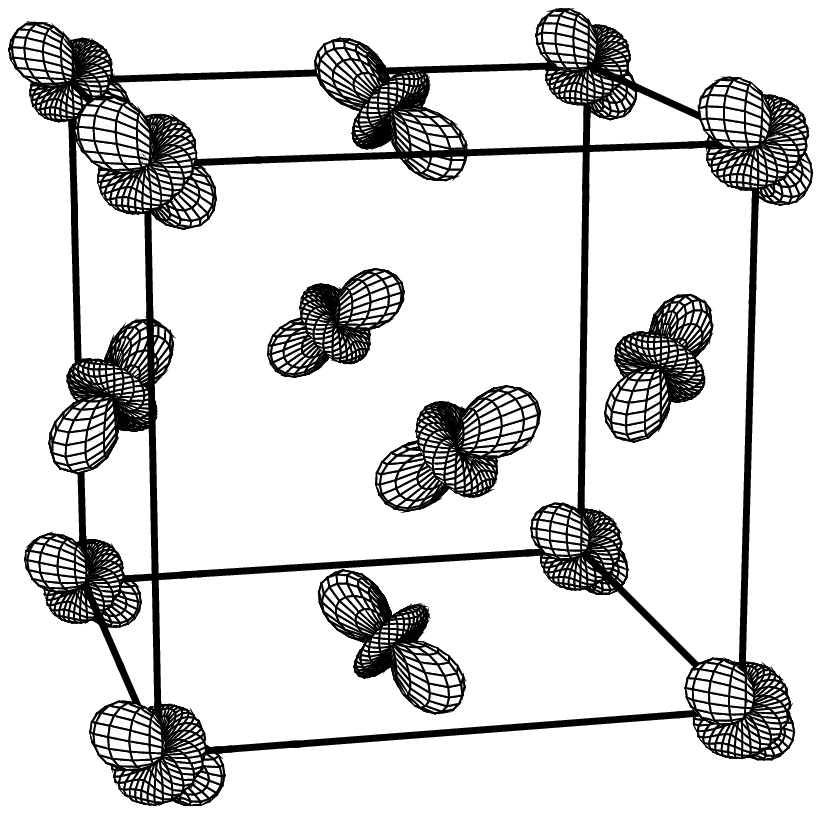}}}
\quad \quad
\subfigure{\resizebox{3.5cm}{!}{\includegraphics{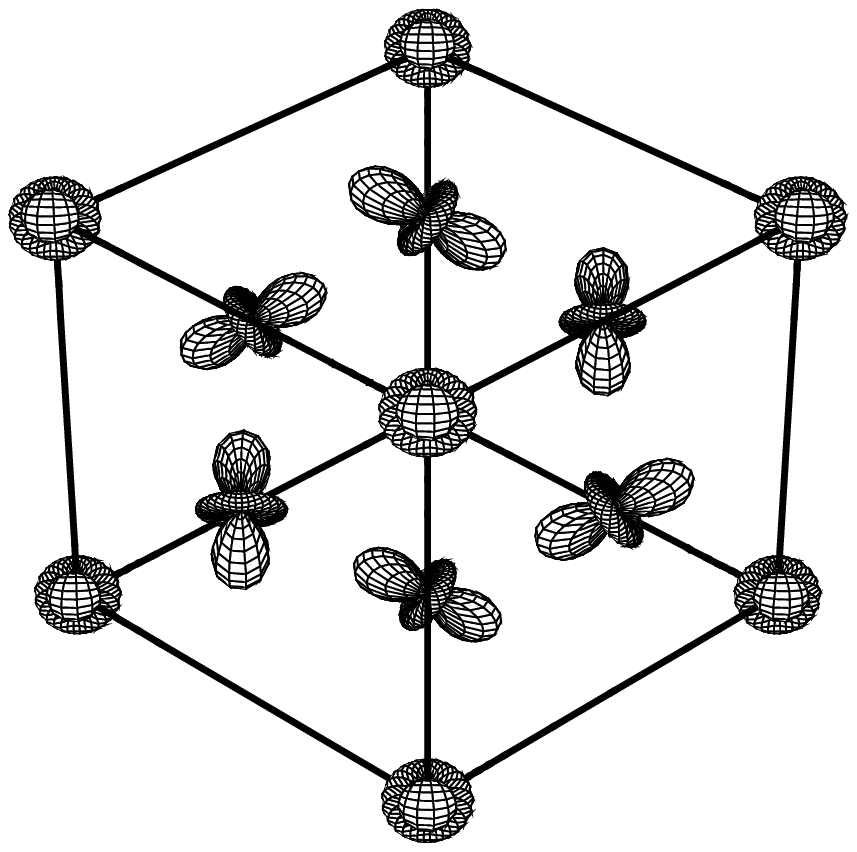}}}
\nonumber

% -----------> Figure Caption
%\vspace{2mm}
\caption{ Triple-$\vec{q}$ antiferroquadrupolar ($Pa{\bar 3}$)
structure of $\alpha$-Ce proposed by Nikolaev and Michel
\cite{NM1}. Quadrupoles represent the $l = 2$ valence electron
($4f+5d6s^2$) charge density distribution.} \label{fig1_TDPAC}
\end{figure}

On the other hand, on the basis of thermal expansion and compressibility data it was suggested \cite{EC} that in Ce instead of the critical point \cite{Kos,Lip}
a tricritical point \cite{Lan1} is realized. In such a case a first order phase transition between two phases of different space symmetry continues beyond the tricritical point as a second order phase transition \cite{EC}. Therefore, $\alpha$-Ce should have a lower symmetry than $\gamma$-Ce \cite{EC}. A mechanism of symmetry lowering was provided by a new scenario of quadrupolar electron charge density ordering \cite{NM1,NM2}, which contrary to the previous approaches predicts a hidden and very peculiar symmetry change at the $\gamma \rightarrow \alpha$ transition, Fig.~\ref{fig1_TDPAC}.
%%%%%%%%%%%%%%%%%%%%%%%%%%%%%%%%%%%%%%%%%%%%%%%%%%%%%%%%%%%%%%%%%%
%
%------------------------------------------------------
%    FIGURE 1
%------------------------------------------------------
%
%%%%%%%%%%%%%%%%%%%%%%%%%%%%%%%%%%%%%%%%%%%%%%%%%%%%%%%%%%%%%%%%%
In the present work we present experimental evidence for the presence of local quadrupolar electronic densities in the $\alpha$ phase and the absence in the $\gamma$ phase.
The transition $\gamma \rightarrow \alpha$ is of the first order and is driven by the minimization of the electron repulsion between the neighboring
cerium sites. The active electronic mode belongs to the $X$ point
of the Brillouin zone and involves its three arms,
$\vec{q}_X^{\;x}$, $\vec{q}_X^{\;y}$, $\vec{q}_X^{\;z}$, that
causes the quadrupolar order parameter components alternate sign
along $x$, $y$, and $z$-axis. The space group symmetry lowering, $Fm{\bar
3}m$($\gamma$-Ce) $\rightarrow$ $Pa{\bar 3}$ ($\alpha$-Ce),
is accompanied by a uniform lattice contraction so that the fcc structure
of the atomic centers of mass (cerium nuclei) is fully
conserved \cite{NM1,NM2}. The long range order is due to the orientation of the local $Y_{\ell=2}^{m=0}$ quadrupolar charge density component of the
valence electrons of cerium ($4f+5d6s^2$) on four different sublattices, Fig.~\ref{fig1_TDPAC}. One can speak of a triple-$\vec{q}$ antiferroquadrupole (3-q-AFQ) structure.
The change can not be captured by usual x-ray diffraction
experiments. Such a scenario is in agreement with previous
experimental results, and in addition reconciles the $\gamma -
\alpha$ transformation with the Landau theory of
phase transitions. The significance of the latter fact should not
be underestimated. In particular, it implies that apart from the
specific symmetry change the $\gamma - \alpha$ phase transition is
not fundamentally different from the other cerium transformations: $\alpha
\rightarrow \alpha''$(monoclinic $C2/m$ symmetry), $\alpha''
\rightarrow \alpha'$(orthorhombic $\alpha$-U space symmetry), etc.
\cite{Kos,McMah}. It is also clear that the other theoretical
models for the $\gamma \rightarrow \alpha$ phase transition treat
it as an exclusive case which is not immediately applicable to
other phase transformations. Unlike these other approaches, the
theory of quadrupole ordering \cite{NM1,NM2} appealed to the
experimental refinement of the electronic structure of
$\alpha$-Ce.

The phase transition of the $Fm{\bar 3}m \rightarrow$ $Pa{\bar 3}$
type is not uncommon in molecular crystals. It occurs in fullerite
C$_{60}$ \cite{C60}, nitrogen crystal ($\alpha$-N$_2$) \cite{N2}
etc. A hidden $Fm{\bar 3}m \rightarrow$ $Pn{\bar 3}m$ symmetry change was found experimentally in NpO$_2$ \cite{Pai}. The $Pn{\bar 3}m$ symmetry \cite{Pn3m} there is another triple-$\vec{q}$ structure, which differs from $Pa{\bar 3}$ in the way the threefold axes and electronic quadrupoles are distributed over four sublattices \cite{Nik1}. The problem with $\alpha$-Ce is that the proposed space symmetry does not alter the fcc structure of the cerium nuclei. Although in principle the change can be detected by synchrotron radiation experiments, in practice there are many serious technical problems, which so far have obstructed the experiment.

The existence of the quadrupolar order in $\alpha$-Ce and its site symmetry can be tested by other experimental techniques. In particular, methods of nuclear physics are often used to describe the crystal electric field potential $V$ and the local site symmetry by measuring the electric field gradient (EFG) at nuclear positions: $V_{ij} = \partial ^2 V/\partial i \partial j$ ($i,j=x,y,z$). Being a symmetric traceless second-rank tensor, $V_{ij}$ can be diagonalized. Then it is fully determined by the largest principal component, $V_{zz}$, and the asymmetry parameter $\eta =(V_{xx}-V_{yy})/V_{zz}$ ($0 \le \eta \le 1$) \cite{Ste}. The electric field gradient is proportional to the nuclear quadupole frequency (QF) $\nu_Q = e Q V_{zz}/h$ ($Q$ is the nuclear quadrupole moment), which is customary used for characterization of its strength. Through the electric quadrupole hyperfine interaction the electric field gradient at a lattice site is directly experienced by a probe nucleus. Such nuclear quadrupole interactions in solids are exploited by many famous methods (for example, by nuclear quadrupole resonance, M\"ossbauer spectroscopy, etc.). Below we use a less used technique of time-differential perturbed angular correlations \cite{Ste} (TDPAC) with the $^{111}$In/$^{111}$Cd probe atoms implanted in the cerium lattice \cite{Tsv1,Tsv}. It is worth noting that the TDPAC measurements can be performed with very few impurity nuclei and in contrast to the M\"ossbauer spectroscopy it is not restricted to low temperatures. In the best cases the method yields an accuracy approaching that of nuclear quadupole resonance.

The method exploits an anisotropic intensity distribution of $\gamma$ rays emitted from an ensemble of the (intermediate) excited nuclear states of probe $^{111}$Cd atoms (171-245 keV cascade). The cascade proceeds via the 245 keV level with the half-life $T_{1/2}$ = 84 ns, spin $I = 5/2$ and the nuclear quadrupole moment $Q$ = 0.83 b. The precession of nuclear spins in a static electric crystal field changes the alignment of the ensemble with time, which is observed by measuring the $\gamma$ ray anisotropy [see Eq.~(\ref{e3}) below]. The initial alignment is prepared by populating the excited states with the electron capture decay of the 2.8 d $^{111}$In isotope. The $^{111}$In activity was produced via the $^{109}$Ag $(\alpha,2n)$ $^{111}$In reaction by irradiating a silver foil with the 32 MeV $\alpha$-beam. After that the nuclear
$^{111}$In/$^{111}$Cd probes were implanted in the cerium lattice
by melting Ce powder (about 500 mg) with a small piece of the
irradiated silver foil ( $\le$0.1 mg) in a special chamber under
pressure of 8 GPa \cite{Tsv1}.
The TDPAC measurements were performed on polycrystalline samples of cerium metal at room temperature using a four-detector spectrometer equipped with a
small size hydraulic four arm press with a capacity up to 300 ton
\cite{Bru}. The high pressure up to 8 GPa was produced in a
calibrated ``toroid"-type device with NaCl as pressure-transmitting medium, Ref.~\cite{Tsv}. Non-hydrostaticity of the transmitting medium was checked by measuring the $^{111}$Cd-TDPAC spectra of silver (fcc) under high pressure and observed to be negligibly small.

The angular time correlation is characterized by the second order
perturbation factor of the static electric quadrupole interaction (QI) for the nuclear spin $I = 5/2$, which describes its precession due to a hyperfine
interaction \cite{Ste}:
\begin{eqnarray}
     G_{22}(t;\; \nu_Q,\, \eta,\, \Lambda) = s_{20} + \sum_{n=1}^3 s_{2n}\,
     \cos(\omega_n t)\, e^{-\Lambda_n \omega_n t} .
\label{e2}
\end{eqnarray}
Here $t$ is the time delay, $s_{2n}$ are the amplitude
coefficients, $\omega_n$ are the angular precession frequencies
($n = 1,2,3$ for $I = 5/2$) related to the energy differences
between hyperfine levels split by the QI. The frequencies are
functions of the EFG $V_{zz}$ or, alternatively, $\nu_Q$, and the asymmetry parameter $\eta$. $\Lambda$ is the half-width of the QF Lorentzian distribution accounting for oscillatory damping.
%%%%%%%%%%%%%%%%%%%%%%%%%%%%%%%%%%%%%%%%%%%%%%%%%%%%%%%%%%%%%%%%%%
%
%------------------------------------------------------
%    FIGURE 2
%------------------------------------------------------
\begin{figure*}
\subfigure{\resizebox{5.3cm}{!}{\includegraphics{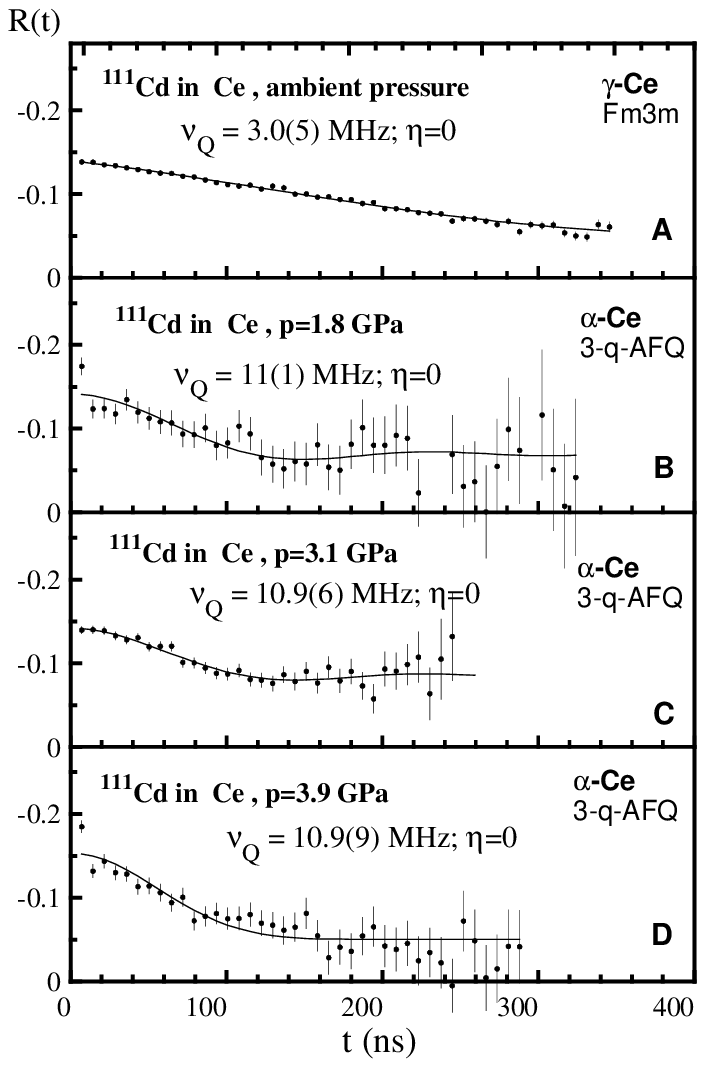}}}
\quad \quad \quad
\subfigure{\resizebox{5.3cm}{!}{\includegraphics{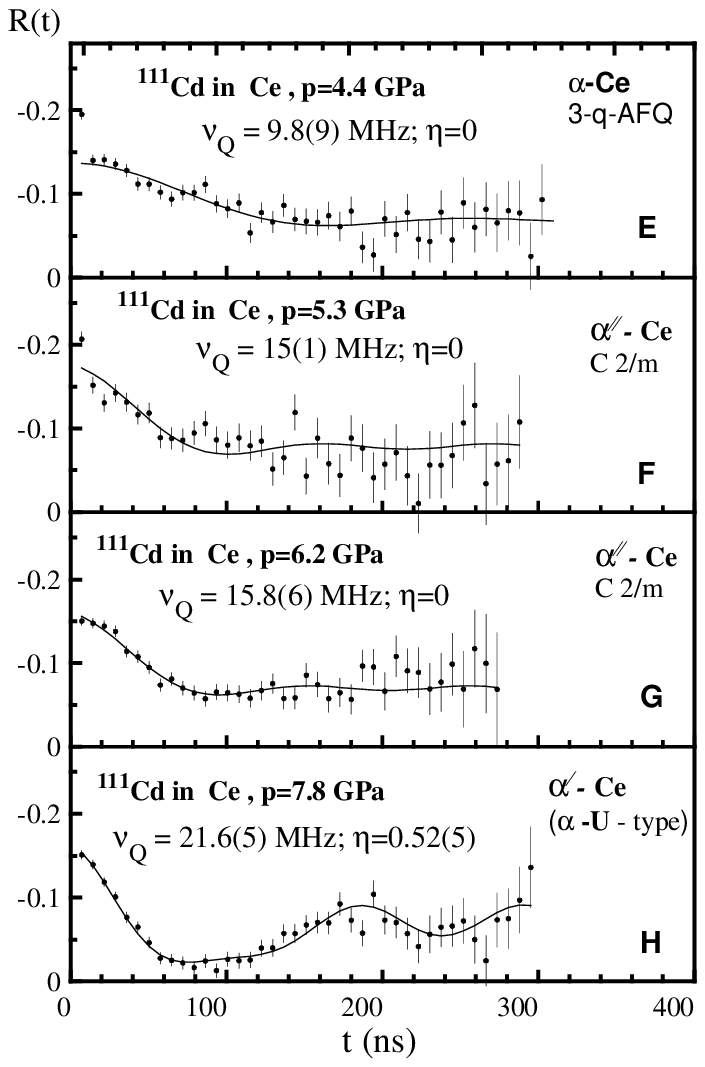}}}
\nonumber

\caption{ Room temperature TDPAC spectra of $^{111}$Cd in cerium
under pressure. (A) at normal pressure: $\gamma$-Ce, fcc [$Fm{\bar
3}m$] structure, no quadrupole order; (B), (C), (D), (E)
$\alpha$-Ce, 3-q-AFQ structure at 1.8, 3.1, 3.9, and 4.4 GPa,
respectively. (F), (G) $\alpha''$-Ce [$C2/m$] at 5.3 and 6.2 GPa;
(H) $\alpha'$-Ce [$\alpha$-U-structure] at 7.8 GPa.}
\label{fig2_TDPAC}

\end{figure*}
%
%%%%%%%%%%%%%%%%%%%%%%%%%%%%%%%%%%%%%%%%%%%%%%%%%%%%%%%%%%%%%%%%%

Recording the delayed coincidence spectra at angles $\pi/2$ and
$\pi$ between detectors, $N(\pi/2,\, t)$ and $N(\pi,\, t)$, one
obtains the usual angular anisotropy
\begin{eqnarray}
   R(t) = -2\frac{N(\pi,\, t) - N(\pi/2,\, t)} {N(\pi,\, t) + 2N(\pi/2,\, t)} ,
\label{e3}
\end{eqnarray}
which is the TDPAC spectrum. The pressure evolution of the TDPAC
spectrum of $^{111}$Cd in Ce is given in Fig.~\ref{fig2_TDPAC}. Experiments have been carried out at room temperature and covered four phases of elemental cerium: $\gamma$, $\alpha$, $\alpha''$ and $\alpha'$. It
can be shown \cite{Ste} that $R(t) = -A_{22} Q_2 G_{22}(t)$, where
$G_{22}(t)$ is the perturbation factor, Eq.~(\ref{e2}), $Q_2
\approx$0.80 is the solid-angle correction and $A_{22} = -0.17$ is
the unperturbed angular correlation coefficient for the
$\gamma-\gamma$ cascade of $^{111}$Cd. The EFG parameters are
determined from a least-square fitting of the TDPAC spectra in
accordance with Eq.~(\ref{e2}).

The extensive TDPAC data are presently available for the
$^{111}$Cd impurity, for which the EFG as a function of
temperature and pressure has been determined in several rare earth
(RE) metals \cite{For1,Via}. It should be noted that the majority
of RE metals crystallizes in hexagonal structures implying a
nonzero EFG. In particular, the double hexagonal
close packed (dhcp) $\beta$ phase of cerium has been thoroughly
investigated by Forker {\it et al.}~\cite{For}.

\begin{figure}
\vspace{1mm} \resizebox{0.4\textwidth}{!} {
 \includegraphics{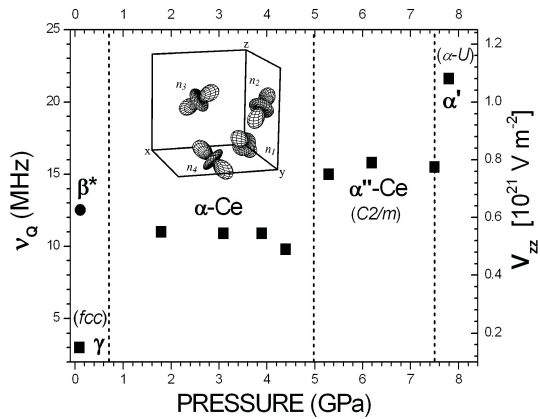}
}
% -----------> Figure Caption
%\vspace{2mm}
\caption{ Pressure dependence of EFG ($V_{zz}$, right scale) and
the nuclear QI frequency ($\nu_Q$, left scale) of $^{111}$Cd
implanted in cerium. $^*$Data for $\beta$-Ce are from Forker {\it
et al.}, Ref.~\cite{For}. } \label{fig3_TDPAC}
\end{figure}

In the cubic symmetry the first nontrivial electron charge density
contribution at each site is described by the cubic harmonic $K_4$
(i.e. $l=4$) thus leaving no room for an EFG tensor: $V_{ij} = 0$.
Therefore, one should find $\nu_Q \approx 0$ and $V_{ij} \approx
0$ if both $\gamma$ and $\alpha$-Ce are assumed to be
isostructural. Instead, from our data for cubic phases, Figs.~\ref{fig2_TDPAC} and \ref{fig3_TDPAC}, we find that the conclusion holds only for $\gamma$-Ce, while in $\alpha$-Ce $\nu_Q
\ne 0$ and $V_{ij} \ne 0$ signaling the appearance of a hidden
quadrupole charge density component at the probe $^{111}$Cd atoms.

We first describe the TDPAC spectrum of $\gamma$-Ce,
Fig.~\ref{fig2_TDPAC}(A). At ambient pressure and room temperature the fitting gives $\nu_Q =$3.0(5) MHz ($\eta=0$). Notice, that our TDPAC spectrum of $\gamma$-Ce is very close to the one observed by Forker {\it et al.} (upper panel of Fig.~1 of Ref.~\cite{For}) with the same value of $\nu_Q$. A small electric field gradient attributed to lattice defects and other impurities is found in all cubic metals and merely reflects the purity of the material.

From the structural data we know that the TDPAC spectra at
pressures 1.8, 3.1, 3.9, and 4.4 GPa should be identified as
belonging to the $\alpha$-phase, Figs.~\ref{fig2_TDPAC}(B)-(E). They are consistent and from fitting we obtain $\nu_Q = 11(1)$ MHz
($V_{zz}=$0.54(4) $\times$10$^{21}$V$\cdot$m$^{-2}$) and a
uniaxial local site symmetry ($\eta = 0$), Fig.\ \ref{fig3_TDPAC}.
Furthermore, the values of $\nu_Q$ and $V_{zz}$ are close to the TDPAC
parameters for $\beta$-Ce ($\nu_Q =$12.5(7) MHz,
$V_{zz}=$0.61(4)$\times$10$^{21}$V$\cdot$m$^{-2}$ \cite{For}), and
other noncubic phases of cerium, Fig.\ \ref{fig3_TDPAC}. The observed values of QF and EFG for $\alpha$-Ce unambiguously rule out the cubic symmetry and indicate a hidden quadrupolar order in this phase. This experimental finding can be rationalized only by assuming a 3-q-AFQ order in $\alpha$-Ce,
Fig.~\ref{fig1_TDPAC}. The only theory predicting such a symmetry
change for cerium was developed by Nikolaev and Michel
\cite{NM1,NM2}.

The $^{111}$Cd TDPAC spectra at pressures 5 to 8 GPa are shown in
Fig.~\ref{fig2_TDPAC}(F)-(H). They show two step-wise changes characteristic of two other phase transformations. An increase in the quadrupole frequency at 5.3 GPa without the change of the asymmetry parameter ($\eta = 0$) is induced by the modification of the Ce crystal structure from the simple cubic symmetry ($\alpha$-Ce) to the monoclinic $\alpha''$ phase. A partial ``softening" of the EFG and $\nu_Q$ in $\alpha$-Ce at 4.4 GPa is also an indication of a proximity to the phase boundary. According to the angle-dispersive powder diffraction technique \cite{McMah} the space symmetry of $\alpha''$-Ce is $C2/m$. From Fig.~\ref{fig2_TDPAC}, \ref{fig3_TDPAC} one sees that the monoclinic $C2/m$ structure is preserved up to 7.5 GPa. At pressure 7.8 GPa the $^{111}$Cd TDPAC spectrum characteristics change again. Now, it is the orthorhombic ($\alpha$-U) structure \cite{McMah}. The assignment is corroborated by a large value of $\eta=0.52$, typical for the $\alpha$-U symmetry. (For elemental uranium in the $\alpha$-phase $\eta=1$~\cite{Hut}.)

Fig.~\ref{fig3_TDPAC} presents the overall pressure dependence of
QF and EFG in Ce. At the phase transition boundaries QF and EFG
change discontinuously indicating the first-order character of the
transitions. Earlier, the step-wise behavior of QF with
changes of the crystal structure was observed for a number of rare
earths \cite{For1,Via}.
%%%%%%%%%%%%%%%%%%%%%%%%%%%%%%%%%%%%%%%%%%%%%%%%%%%%%%%%%%%%%%%%%%
%
%------------------------------------------------------
%    FIGURE 3
%------------------------------------------------------
%
%%%%%%%%%%%%%%%%%%%%%%%%%%%%%%%%%%%%%%%%%%%%%%%%%%%%%%%%%%%%%%%%%

In conclusion, our TDPAC experiments (Fig.~\ref{fig3_TDPAC})
detect an appreciable EFG in $\alpha$-Ce comparable with EFG for
noncubic phases ($\beta$, $\alpha''$), which border $\alpha$-Ce in
the pressure-temperature phase diagram. This finding rules out the
cubic symmetry in $\alpha$-Ce and evidences in support of a
3-q-AFQ order suggested by Nikolaev and Michel \cite{NM1}. There,
the local site symmetry is trigonal ($C_3$) with a quadrupolar
electron charge density component $Y_{\ell=2}^{m=0}$ oriented
along one of the main cube diagonals, Fig.~\ref{fig1_TDPAC}. With
further pressure increase, the evolution of the TDPAC spectrum
follows other phase transitions in cerium: $\alpha \rightarrow
\alpha''$ ($C2/m$) and $\alpha'' \rightarrow \alpha'$ ($\alpha$-U
structure).

This work is supported by the Program of the Presidium of
RAS ``Physics of Strongly Compressed Matter". We are grateful to
Prof. S.M. Stishov, Prof. K.H. Michel, Dr. B. Verberck, Prof.
A.N. Grum-Grzhimailo and Prof. V.B. Brudanin for support of this work and discussion of the results.

%---------------- REFERENCES -------------------------------


\begin{thebibliography}{99}
\bibitem{Pn3m}
$Pa{\bar 3}$ minimizes the electron-electron repulsion on
neighboring sites, $Pn{\bar 3}m$ can win as a result of
$f$-electron participation in metal bond formation.
\end{thebibliography}
\end{document}